\documentstyle[epsfig,referee,a4]{l-aa}

\newcommand{\gccm}{~\rm{g~cm}^{-3}}
\newcommand{\sfs}{$^1\rm{S}_0$}
\newcommand{\sfp}{$^3\rm{P}_2$}

\begin{document}

%%%%%%%%%%%%%%%%%% TITLE AND AUTHOR INFORMATION %%%%%%%%%%%%%%%%%%%%%%%%%%%%%
\thesaurus{08(     % Stars
	08.14.1;   % Stars: neutron
	08.05.3;   % Stars: evolution
	02.04.1;   % dense matter
	13.25.5)}   % X-rays: stars

\title{Impact of medium effects on the cooling of non--superfluid and
  superfluid neutron stars}

\author{Ch. Schaab\inst{1} 
        \and D. Voskresensky\inst{2}$^,$\inst{3}
        \and A. D. Sedrakian\inst{4}
	\and F. Weber\inst{1}$^,$\inst{5}
	\and M. K. Weigel\inst{1}}
\offprints{Ch. Schaab, email: schaab@gsm.sue.physik.uni-muenchen.de}
\institute{
	Institute for Theoretical Physics, Ludwig-Maximilians Universit{\"a}t,
        Theresienstrasse 37/III, D-80333 Munich, Germany
\and    Gesellschaft f\"ur Schwerionenforschung GSI, 
        P. O. Box 110552, D-64220 Darmstadt, Germany
\and		Moscow Institute for Physics and Engineering, 
        	115409 Moscow, Kashirskoe shosse 31, Russia
\and    Max--Planck--Society,
      Research unit 'Theoretical Many--Particle Physics' at Rostock
      University, Universit\"atplatz 1, D--18051 Rostock, Germany
\and	    Nuclear Science Division, Lawrence Berkeley National
      	    Laboratory, MS: 70A-3307, University of California, Berkeley, CA
      	    94720, USA}

\date{Received .........; accepted .........}

\maketitle
\markboth{Ch. Schaab et al.: Impact of medium effects on the cooling of 
          neutron stars}{}

%%%%%%%%%%%%%%%%%%%%%%%%% ABSTRACT %%%%%%%%%%%%%%%%%%%%%%%%%%%%%%%%%%%
\begin{abstract}
  Neutrino emission from the dense hadronic component in neutron stars
  is subject to strong modifications due to collective effects in the
  nuclear medium. We implement new estimates of the neutrino
  emissivities of two processes operating in the nuclear medium into
  numerical cooling simulations of neutron stars. The first process is
  the modified Urca process, for which the softening of the pion
  exchange mode and other polarization effects as well as the neutrino
  emission arising from the intermediate reaction states are taken
  into account. The second process concerns neutrino emission through
  superfluid pair breaking and formation processes. It is found that
  the medium effects on the emissivity of the modified Urca process
  result in a strong density dependence, which gives a smooth
  crossover from the standard to the nonstandard cooling scenario for
  increasing star masses. For superfluid stars, the superfluid pair
  breaking and formation processes accelerate mildly both the standard
  and the nonstandard cooling scenario. This leads to a good agreement
  between the theoretical cooling tracks and the rather low
  temperatures observed for objects like PSRs 0833-45 (Vela), 0656+14,
  and 0630+18 (Geminga).  The robustness of our findings against
  variations in both the underlying equation of state of baryonic
  matter and the used fast cooling processes is demonstrated.  Hence
  we conclude that the two recalculated neutrino emissivities studied
  here enable one to reproduce theoretically most of the observed
  pulsar temperatures by varying the masses of neutron star models.
\end{abstract}
        
%%%%%%%%%%%%%%%%%%%%%%%% INTRODUCTION %%%%%%%%%%%
\section{Introduction}\label{sec:intro}

The physics of neutron star cooling is based on a number of
ingredients, among which the neutrino emissivity of the high--density
hadronic matter in the star's core plays an important role.  Depending
on the dominant neutrino--emission process at the early stages of the
thermal evolution, the cooling simulations follow either the slow
(standard) or the fast (nonstandard) scenario of thermal evolution. In
the first case the dominant neutrino--radiation reactions are the
modified Urca and the bremsstrahlung processes. In the second case,
these are the pion (kaon) $\beta$--decay processes, direct Urca on
nucleons and hyperons, as well as their analogous reactions taking
place in the deconfined quark phase.  The main difference in the
cooling efficiency driven by these processes lies in the rather
different phase spaces associated with these reactions.  In the first
case the available phase space is that of a two--baryon scattering
process, while in the second case it is that of a one--baryon decay
process.

Numerical simulations of neutron star cooling, incorporating these
types of neutrino emission, have been extensively performed in the
past (see, for instance, Tsuruta \cite{Tsuruta66}, Richardson et al.
\cite{Richardson82}, Van Riper \cite{VanRiper91}, and Schaab et al. 
\cite{Schaab95a}).  These 
calculations suggest that the combined soft X-ray data from Einstein,
EXO\-SAT, and ROSAT are roughly consistent with the slow cooling
scenario, depending on the equation of state and the uncertainties
associated with the behavior of super--dense matter.  Some pulsars,
however, as for example Vela and Geminga, possess rather low
temperatures and thus seem to call for a more rapid cooling than is
obtained for the modified Urca process.  The fast cooling scenarios,
on the other hand, generally have the tendency to underestimate the
surface temperatures of these pulsars. So a natural question,
motivated by these observations, is whether the physics of neutrino
emission from the star's dense core will be able to resolve this
problem. This will be discussed in this paper. Obviously, the
corresponding modifications must be such that the theoretical cooling
curves describe both the hotter as well as cooler classes of pulsars,
which could be linked to variations in gross--structure parameters,
like the star's mass (Voskresensky \& Senatorov 1984
(\cite{Voskresenskii84}), 1986 (\cite{Voskresenskii86})).

The aim of the present work is to implement a number of
medium--modified neutrino emissivities, which dominate a neutron
star's thermal evolution, in numerical cooling simulations of such
objects.  We shall demonstrate that by means of this one is indeed
able to achieve agreement with both the high as well as low observed
pulsar temperatures.

It is a well established fact that neutrino emission at the early
stages of the standard evolution is dominated by the modified Urca and
bremsstrahlung processes. The neutrino radiation in the modified Urca
process was first estimated by Bahcall \& Wolf (\cite{Bahcall65}) and
by Tsuruta \& Cameron (\cite{Tsuruta65}).  Bremsstrahlung processes
dealing with the neutral currents were first evaluated by Flowers et
al. (\cite{Flowers75}).  Then the corresponding rates have been more
closely studied by Friman \& Maxwell (\cite{Friman79},FM79) resulting
in sufficiently higher rates. The calculation were performed in the free
one--pion--exchange approximation to the long--range part of the
nucleon--nucleon (NN) interaction, supplemented by a parametrization
of the short--range part of the NN interaction by the Landau
Fermi--liquid parameters.  Medium effects enter these rates mainly
through the effective mass of the nucleons. Therefore the density
dependence of the rates is rather weak and the neutrino radiation from
a neutron star depends only very weakly on its mass.  This is the
reason why the standard scenario based on the FM79 result, though
complying well with a few slowly cooling pulsars, fails to explain the
data of the more rapidly cooling ones.

Here, we shall carry out detailed simulations of the cooling of
neutron star models, which incorporate the softening of the one--pion
exchange mode, other medium polarization effects, like the inclusion
of the nucleon-nucleon correlations in the vertices, as well as
the possibility of neutrino emission from intermediate reaction
states\footnote{E.g., the pion participating in the exchange between
the nucleons may also radiate the neutrinos or decay in intermediate
reaction states to the nucleon-nucleon hole with subsequent neutrino
radiation. This essentially modifies the absolute value as well as the
density dependence of the modified Urca process rate.}
(\cite{Voskresenskii84,Voskresenskii86}). We also include the
processes of neutrino pair radiation from superfluid nucleon pair
breaking and formation mechanism first estimated by Flowers et al.
(\cite{Flowers76b}), and then more closely studied by
Voskresensky
\& Senatorov (\cite{Voskresenskii87a}) and Senatorov \& Voskresensky
(\cite{Senatorov87}) using the closed diagram technique. The medium 
modifications of these rates result in a pronounced density
dependence, which, in turn, links the cooling behavior of a neutron
star decisively to its mass.

Using a collection of modern equations of state for nuclear matter in
its ground state, which covers both relativistic as well as
non--relativistic models, we shall demonstrate the robustness of the
new cooling mechanisms against variations in the equation of state of
super-dense neutron star matter.  Part of these variations are caused
by the possible superfluid behavior of neutron star matter. To
demonstrate this effect on the new cooling mechanisms, we proceed in
three successive steps, starting from stars made up of non-superfluid
matter. These models are then supplemented with superfluidity, with
superfluid gap parameters taken from the literature. Unfortunately,
the gap energies are not very well known because of uncertainties in
the effective baryon masses, the nucleon--nucleon interaction, and
polarization effects (cf. Chen et al. \cite{Chen93}). To account for
these uncertainties, their values are finally varied about the
theoretically determined ones. Since there are also uncertainties
concerning the possibility of fast cooling processes the calculations
are done with both the nucleon direct Urca and the $\beta$-decay
processes in pion condensates (pion Urca) and additionally without any
fast cooling process.

We shall show that the above mentioned medium--modifications lead to a
more rapid cooling than obtained for the standard scenario. Hence they
provide a possible explanation for the observed deviations of some of
the pulsar temperatures from standard cooling.  Particularly, it is
shown that they provide a smooth transition from standard to
nonstandard cooling for increasing central star densities, i.e., star
masses.

The paper is organized as follows.  In Sect. 2 we briefly discuss the
incorporated neutrino emission processes which comprise the modified
Urca, superfluid nucleon--pair breaking and formation, direct Urca and
pion Urca processes. In Sect. 3 the neutron star cooling simulations
and the physical input quantities used in the calculations are
described.  In Sect. 4 we present the results of the numerical
calculations and compare them with the observed data.  The conclusions
are summarized in the last section.

%%%%%%%%%%% EMISSIVITIES OF THE MAIN PROCESSES %%%%%%%%%%%%%%%%%%%%%%%%%%%%%%
\section{Neutrino Emissivities} 

\subsection{Medium Effects on the Modified Urca Process} \label{sec:emis.smu}

\begin{figure}
%%\picplace{4cm}
\centering\psfig{figure=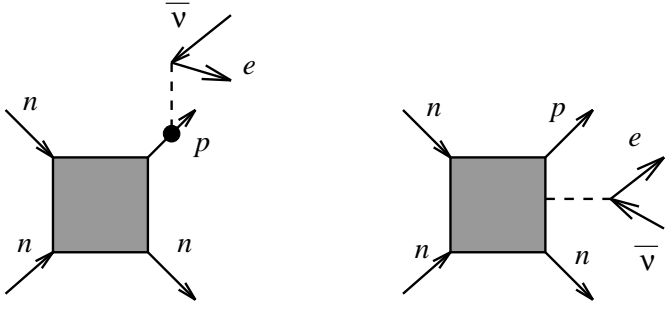,width=8.8cm}
\caption[]{Neutrino emission from a nucleon leg (left graph) and 
  from intermediate scattering states (right graph). Details are given
  in the text.
 \label{fig:graph1}}
\end{figure}

The modified Urca (MU\footnote{We shall abbreviate the modified Urca
process calculated via the free one--pion exchange model by Friman and
Maxwell with MU-FM79. The abbreviation MU-VS86 will refer to the process
which accounts for the additional included medium modification
effects. The abbreviations of the various neutrino-emitting processes
considered in this paper are summarized in Table \ref{tab:abbrev}.})
process is linked to neutrino emission via a weak decay of a nucleon
in the presence of a bystander nucleon. The latter ensures that both
momentum and energy are conserved in the reaction.  The two
topologically different diagrams contributing to these processes are
schematically shown in Fig. \ref{fig:graph1}. The first diagram shows
the neutrino radiation from a nucleon ``leg'', i.e., from an initial--
or final--state nucleon.  The shaded block represents the total
nucleon--nucleon interaction amplitude in the medium (including the
softening of the pion mode, corrections of the vertices by the
nucleon--nucleon--hole and delta--nucleon--hole loops, and
modifications of the short--range interaction by nucleon-nucleon--hole
and delta--nucleon--hole loops), while the filled circle stands for
the nucleon--nucleon correlations.  The second diagram illustrates the
neutrino radiation from the intermediate scattering states (like
neutrino radiation by the pion and the nucleon-nucleon hole loop being
present in intermediate reaction states). Such medium effects as well
as the second reaction channel were not incorporated in the
calculations by FM79 and shall lead to a rather new cooling behavior,
as we will see below.

\begin{table}
\caption[]{Key to abbreviations \label{tab:abbrev}}
\begin{tabular}{ll}
\hline\noalign{\smallskip}
abbreviation & meaning \\
\noalign{\smallskip}\hline\noalign{\smallskip}
MU-FM79               &modified Urca process calculated by FM79\\
MU-VS86               &modified Urca process calculated by VS86\\
NPBF                  &neutron pair-breaking and formation\\
PPBF                  &proton pair-breaking and formation\\
DU                    &direct Urca process \\
PU                    &pion Urca process\\
\noalign{\smallskip}\hline
\end{tabular}
\end{table}
 
The main contribution to the nucleon--nucleon interaction amplitude at
densities $n\ga 0.5-0.7\, n_0$, where $n_0 \simeq 0.16$ fm$^{-3}$ is
the nuclear matter saturation density, is determined by the in--medium
one--pion--exchange supplemented by medium--modified vertices
(\cite{Voskresenskii86}), whereas the peculiarities of a more local
part of the interaction play an important part only at rather small
densities, $n\la 0.5n_0$ (Blaschke et al. \cite{Blaschke95}). At $n\ga
n_0$ the emissivity arising from the modified Urca is dominated by the
second diagram of Fig. \ref{fig:graph1}, whereas the first diagram
gives only small corrections (\cite{Voskresenskii86}, Voskresensky et
al. \cite{Voskresenskii87b}, Haubold et al. \cite{Haubold88}). The
contributions of the first diagram, which gives the only contribution
to emissivity calculated by FM79, can therefore be neglected. The
resulting emissivity reads in a simplified notation as follows
(\cite{Voskresenskii86}, Migdal et al. \cite{Migdal90}),
\begin{eqnarray}\label{GU}
  \epsilon_\nu^{(\rm MU-VS86)}&\simeq& 2.4\times 10^{24}~ F_1 \left(\frac{n
    }{n_0}\right)^{10/3} \left(\frac{m^{\ast}_N (n)}{m_N}\right)^4 \nonumber \\
  &\times&  \left[\frac{m_\pi}{ \alpha\, \tilde{\omega}[p_{Fn} (n)]}\right]^8\,
  \Gamma^8~ T^{8}_9 \nonumber \\ &\times&
  \zeta(\Delta_n)~\zeta(\Delta_p)~ {\rm\frac{erg}{cm^3 ~sec}} ~,
\end{eqnarray} 
where $T_9 =T/10^9 K$ is the temperature, $m^*_N$ and $m_N$ are the
effective and bare nucleon mass, respectively, $p_{Fn} (n)$ is the
density dependent neutron Fermi--momentum, and the factor $$F_1
=1+\frac{3}{4\Gamma^2}\, \left(\frac{n}{n_0}\right)^{2/3}$$ is the
correction due to the pion decay from intermediate states. It reduces
to unity when the intermediate state pion decay processes are ignored
and only contributions from the weak current decay in the nucleon
nucleon--hole loop being present in intermediate reaction states are
taken into account.

The quantity $\Gamma$ accounts for the nucleon--nucleon correlations
in the $\pi NN$ vertices.  The effective pion gap, $\tilde \omega^2$,
expressed in units of $m_{\pi} = 140$ MeV, is given by
$\tilde{\omega}[p_{Fn}(n)] \simeq -D^{-1}_\pi\left(\omega\simeq\mu_\pi ,
k=p_{Fn}(n)\right)$, where $D_\pi^{-1}$ is the in-medium pion Green function,
$\mu_\pi$ is the pion chemical potential. (For simplicity, we set the
chemical potentials of $\pi^+ ,
\pi^- $, and $\pi^0$ mesons equal to $\mu_\pi \simeq 0$.)

\begin{figure} 
%\picplace{6.4cm}
\centering\psfig{figure=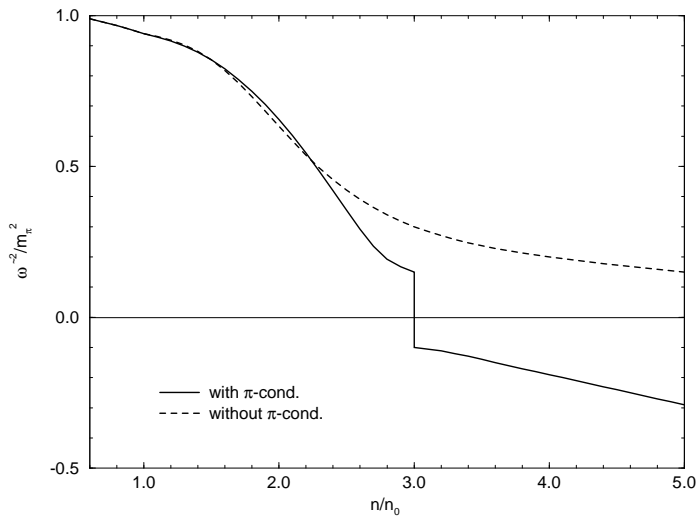,height=8.8cm,angle=-90}
\caption[]{Effective pion gap $\tilde\omega^2/m_\pi^2$ versus
density with (solid line) and without pion condensation (dashed
line). $n_0=$ 0.16~$\rm{fm}^{-3}$ denotes the nuclear saturation
density.
\label{fig:piongap}}
\end{figure}

The values of $\tilde \omega^2$ and $\Gamma$ for isospin symmetric
nuclear matter at saturation density are extracted from the atomic nuclei
data. For the case of interest, $\mu_\pi =0$, they read $\Gamma
(n_0)\simeq 0.4$ and $
\tilde{\omega}^2 (n_0)\simeq 0.8-0.95$ 
(Migdal et al. \cite{Migdal90}).  For highly asymmetric nuclear
matter, especially for $n >n_0$, the magnitude and density dependence
of $\tilde{\omega}^2(n)$ and $\Gamma (n)$ are not well known.  In the
present calculations we approximate the form-factor in the 
effective $\pi NN$ vertex,
$\Gamma(n)$, by the formula
\begin{equation} \label{eq:Gamma}
  \Gamma(n) \approx \frac{1}{1+1.4 \left( n/n_0 \right)^{1/3}}~,
\end{equation} 
which shows the explicit functional dependence on the
density.\footnote{This implies in our case that the Landau-Migdal
parameter $g^{\prime}$ scales as $(n/n_{0})^{1/3}$ leading to an
increasing repulsion for higher densities.}  The effective pion gap
$\tilde{\omega}^2 (n)$ has been estimated in Migdal (\cite{Migdal78}),
Voskresensky \& Mishustin (\cite{Voskresenskii78,Voskresenskii82}),
and Migdal et al. (\cite{Migdal90}) for different parameter
choices. We apply two different parametrizations in the cooling
calculations (see Fig. \ref{fig:piongap}). The first one assumes a
phase transition into the pion condensate phase at $n_c=3n_0$ (Brown
\& Weise \cite{Brown76}, Migdal \cite{Migdal78}, Migdal et
al. \cite{Migdal90}), while the second one assumes that there is no
phase transition (for a more detailed discussion see
Sect. \ref{sec:du}). Due to the pion--fluctuation effect the pion
condensation sets in via a first order phase transition (see Dyugaev
\cite{Dyugaev75}, Voskresensky \& Mishustin
\cite{Voskresenskii78,Voskresenskii82}).  This manifests itself in a
jump of the pion gap at $n=n^{\pi}_c$ from a positive value to a
negative one, see solid line in Fig 2.\footnote{Without this behavior
one would get an anomalous increase of the neutrino emissivity via
bremsstrahlung in the vicinity of the critical point. This effect is
analogous to the critical opalescence phenomenon
(\cite{Voskresenskii86}).}  It has to be emphasized that the density
dependence of $\tilde{\omega}^2 (n)$ is rather unknown. Our choices
for the parametrization are motivated by microscopic many-body
calculations. Somewhat different choices for the values of
$\tilde{\omega}^2 (n)$, $\Gamma (n)$ and the critical density for the
onset of pion condensation affect of course the resulting cooling
rates. The qualitative conclusions, however, remain unchanged. Indeed,
as we will see below, the cooling curves for the two different
parametrizations of $\tilde{\omega}^2 (n)$ do not differ very much
(see e.g. Fig. \ref{fig:hvh_nsf}).

The value of the parameter $\alpha$ is given by $\alpha =1$ for
$n<n^{\pi}_c$, and $\alpha =\sqrt{2}$ for $n>n^{\pi}_c$.  This
parameter accounts for the changes caused by the neutrino emissivity
of the modified Urca process below and above the critical point where
pion condensation occurs (cf. \cite{Voskresenskii86}).

The factor 
\begin{eqnarray}
  \zeta(\Delta_{n}) = \left\{ \begin{array}{ll} {\rm
    exp}\left(-\Delta_n/k_{\rm B}T\right) & \mbox{ $ T \le T_{cn}$},\\ 1 &
  \mbox{$T > T_{cn}$}
\end{array}
\right.
\end{eqnarray} 
takes roughly into account the suppression caused by the nucleon
pairing, where $\Delta_n$ and $T_{cn}$ are the superfluid gap and the
critical temperature for superfluid phase transition of neutrons. The
factor $\zeta(\Delta_p)$ is defined in the analogous way and takes
into account the proton--proton pairing.

%...................................................................
\subsection{Cooper Pair--Breaking and Pair--Formation Processes}
\label{ssec:pair}

When a neutron star cools down to temperatures in the vicinity of the
critical temperature for pairing of neutrons and protons, pairing
correlations play an increasingly important role in the dynamics of
the star's thermal evolution. Generally, the onset of superfluidity
tends to slow down the cooling rate of a neutron star, since the
neutrino emission processes are drastically suppressed. It enables
however two additional processes as noticed in Flowers et
al. (\cite{Flowers76b}) and Voskresensky \& Senatorov
(\cite{Voskresenskii87a}): the superfluid neutron--pair 
breaking and formation process (NPBF) and the superfluid proton--pair 
breaking and formation process (PPBF).

\begin{figure}
%\picplace{2.9cm}
\centering\psfig{figure=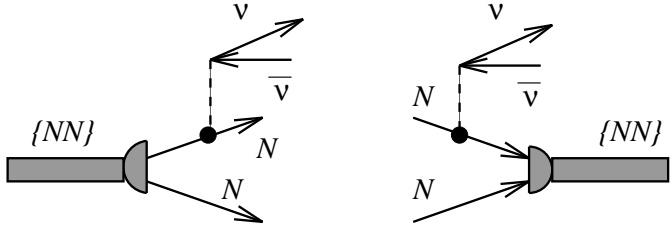,width=8.8cm}
\caption[]{Neutrino emission from Cooper pair-breaking (left graph)
  and pair-formation processes (right graph). Details are given in the
  text.
 \label{fig:graph2}}
\end{figure}

The superfluid in a neutron star can be considered as a two--component
system, which, for a fixed density and temperature, consists of paired
quasiparticles in the condensate and elementary excitations above the
condensate. Their associated quasi--equilibrium densities are
controlled by Cooper--pair--formation and pair--breaking processes.
These processes, which are rather frequent at temperatures in the
vicinity of $T_c$, become successively suppressed at lower
temperatures, because of an exponential increase in the number of
paired particles. As shown in Fig. \ref{fig:graph2}, these
processes proceed with the emission of neutrino pairs via the
reactions $\{NN\}\rightarrow N + N + \nu +\bar{\nu}$ and $N +
N\rightarrow \{NN\} + \nu +\bar{\nu}$, where $\{NN\}$ denotes the
Cooper pair, $N$ an excitation.

The respective emission rates can be calculated either in terms of the
closed diagram formalism, using the normal and anomalous Green's
functions for the paired nucleons, or in terms of the method of
Bogoliubov's transformations applied to reaction rates in the normal
ground state.  As far as one works in the quasiparticle limit, which
is an excellent approximation for neutron star matter, the preference
for one or the other method is primarily a matter of taste. However,
in a broader context, the Green's functions method provides a more
general treatment by allowing for off-mass-shell effects and a
systematic diagrammatical representation of the approximations to the
self-energy functions (scattering rates).  A first estimate of the
reaction rates was given by Flowers et al. (\cite{Flowers76b}) using
the Bogoliubov's transformations. It was recalculated by Voskresensky
\& Senatorov (\cite{Voskresenskii87a}) using closed diagram
technique. Voskresensky \& Senatorov have also included the
contribution of the axial vector coupling and used different
parameterizations of the weak coupling vertices. This results in a
rate being higher by an order of magnitude. Besides the NPBF process,
the PPBF process was considered, too. One finds in lowest order of the
expansion parameters $T/\epsilon_{Fn}$ and $T/\epsilon_{Fp}$
($\epsilon_{Fn}$ and $\epsilon_{Fn}$ are the Fermi--energies of
neutrons and protons, respectively) for the neutron and proton
components (see Voskresensky
\& Senatorov
\cite{Voskresenskii87a})
\begin{eqnarray}\label{NPFB}
  \epsilon_\nu^{(NPBF)}&\simeq& 6.6 \times 10^{28}
  \left(\frac{n}{n_0}\right)^{1/3}\frac{m^{\ast}_N} {m_N}
  \left(\frac{\Delta_n }{{\rm MeV}}\right)^7 \nonumber\\ 
    &\times & I\left(\frac{\Delta_n} 
  {T}\right) \quad \rm{for~} T<T_{\rm c}^{\rm n}~, \\
\label{PPFB} 
\epsilon_\nu^{(PPBF)}&\simeq &1.7 \times 10^{28}
  \left(\frac{n}{n_0}\right)^{1/3}\frac{m^{\ast}_N} {m_N}
  \left(\frac{\Delta_p }{{\rm MeV}}\right)^7 \nonumber\\ 
    &\times & I\left(\frac{\Delta_p}
  {T}\right) \quad \rm{for~} T<T_{\rm c}^{\rm p}~,
\end{eqnarray}
where
\begin{eqnarray}\label{Int}
  I(x) &=& \int^{\infty}_{0} \frac{(\mbox{ch}y)^5
    dy}{[\mbox{exp}(x\mbox{ch}y)+1]^2}  \nonumber\\
  & \simeq & e^{-2x}\sqrt{\pi /4x} \quad {\rm for }\quad x=\Delta /T 
\gg 1 ~.
\end{eqnarray}
It can be seen that in the limit $\Delta /T \gg 1$, the rate of these
processes is exponentially suppressed, as it is the case for the
two--nucleon process considered above. However, because of mild phase
space restrictions (we deal here effectively with a one--nucleon
phase--space volume) these processes may considerably contribute to
the neutrino emissivity of the neutron star. The maximum value of
$\left(\Delta(T)/ \Delta(T=0)\right)^7 I\left(\Delta(T)/T\right)$
depends only on the ratio $\Delta(T=0)/T_c$. For \sfs --pairing the
maximum value is equal to $\approx 10^{-2}$, for \sfp --pairing it
is $\approx 10^{-6}$. This causes the emissivity of the NPBF process
in the core (\sfp --paring) to be smaller by three orders of magnitude
(for equal gap energies) compared to the emissivity in the crust (\sfs
--pairing).

%...................................................................
\subsection{Direct Urca Process and Pion Urca Process}
\label{sec:du}

Along with the two processes described above we include in the
numerical calculations other relevant channels of neutrino emission,
too.  Though we shall see that the processes described above lead to
``{\it intermediate}'' cooling rates, it still appears useful to
distinguish between ``{\it standard cooling}'' and ``{\it
nonstandard}'' or ``{\it fast cooling}'', depending on whether the
predominant process is the modified Urca process or a faster one, as
nucleon or hyperon direct Urca, processes in pion- or kaon-condensed
matter, or those caused by the presence of quarks in dense matter.

As examples of a fast cooling processes, we shall use the nucleon
direct Urca (DU) and the pion Urca (PU) processes. These processes are
rather representative for other fast processes. Indeed, all the fast
cooling processes lead qualitatively to the same temperature
dependence of the emissivity, $\sim T^6$. Moreover the values of the
critical densities beyond which these processes are possible are close
to each other.

The DU processes
\begin{equation}
  \rm{n} \rightarrow \rm{p}+\rm{e}^-+\bar\nu_{\rm e} \quad \rm{and}
  \quad \rm{p}+\rm{e}^- \rightarrow \rm{n}+\nu_{\rm e}
\end{equation}
can only occur if the proton fraction exceeds some critical value
(about 11--13~\% depending on the composition) in order to fulfill
energy and momentum conservation (see Boguta \cite{Boguta81a} and
Lattimer et al. \cite{Lattimer91}). In contrast to nonrelativistic
EOS's (e.g. UV$_{14}$+UVII) which predict proton fractions that lie
below these critical values, relativistic EOS's contain sufficiently
large proton fractions such that the direct Urca process becomes
possible. However, the proton fraction depends crucially on the
symmetry energy which is unfortunately not well known at higher
densities. We shall use the emissivity calculated by Lattimer et
al. (\cite{Lattimer91}), however corrected by the suppression factor
$\Gamma^2$ (see eq. (\ref{eq:Gamma})) accounting for the
nucleon--nucleon correlations in the weak interaction vertex (for
simplicity, we assume the same suppression factor in all vertices,
cf. discussion in Voskresensky \& Senatorov \cite{Voskresenskii87a}):
\begin{eqnarray}\label{eq:DU}
  \epsilon_\nu^{(\rm DU)}&\simeq& 4\times 10^{27}~ \left(\frac{n_{\rm e}
    }{n_0}\right)^{1/3} \left(\frac{m^{\ast}_N}{m_N}\right)^2
    \Gamma^2~ T^{6}_9 \zeta(\Delta_n)~\zeta(\Delta_p)~ 
    {\rm\frac{erg}{cm^3 ~sec}} ~.
\end{eqnarray} 
The impact of the direct Urca process on the cooling will be shown in
Sect. \ref{sec:finres}, where the final results are discussed,
whereas it is neglected in the other cooling curves in order to
demonstrate the efficiency of MU-VS86 and NPBF and PPBF- processes.

Even more uncertain is the possibility of the hyperon direct Urca (see
Prakash et al. \cite{Prakash92}). Whereas the contribution of the
hyperon direct Urca is small compared to the contribution of the
nucleon direct Urca in the non superfluid case, the hyperon direct
Urca might become important if the nucleons form superfluid pairs. If
however the hyperons become also superfluid, then their contribution
is again neglible. The superfluid phase transition of hyperons has
unfortunately not been studied so far since it is rather difficult to
implement superfluidity in relativistic treatments with hyperons. We
therefore neglect this contribution to the neutrino emissivity.

The pion condensation was suggested by Migdal (\cite{Migdal71}),
Scalapino (\cite{Scalapino72}), and Sawyer (\cite{Sawyer72}) and then
considered by many authors (e.g. Brown \& Weise \cite{Brown76}, Migdal
\cite{Migdal78}, and Migdal et al. \cite{Migdal90}). In the last years,
some arguments against the occurrence of pion condensation in neutron stars
were given in the literature (see, for instance, Brown et al. 
\cite{Brown95a}). These arguments are generally based on the rather strong
increase of the Landau-Migdal parameter $g^{\prime}$ with the nucleon
density. However other effects which might be important for the pion
condensation problem (cf. Migdal et al. \cite{Migdal90}) were not
incorporated. Thus one may conclude that the question whether a pion
condensate occurs in neutron stars is still not settled (see Ericson
\& Weise \cite{Ericson88a} and Kunihiro et al. \cite{Kunihiro93a} for
extensive reviews).  We shall therefore use models both with and
without pion condensates applying two parametrizations of
$\tilde{\omega}^2 (n)$ (see Sect. \ref{sec:emis.smu}).

Additionally the neutrino emissivity in kaon condensation, which is
now favored by various authors (e.g. Brown et al. \cite{Brown88a} and
Thorsson et al. \cite{Thorsson95a}) as well as the critical density
for the onset of condensation are quite similar to the case of pion
condensation. The additional $\tan(\Theta_{\rm c})$-factor in the kaon
condensation case, where $\Theta_{\rm c}\simeq0.223$ is the
Cabibbo--angle, is partly compensated by a probably larger value of
the suppression-factor $\Gamma^4$ due to a possible suppression of the
correlations between strange and non-strange baryons. The cooling
curves for the models with kaon condensation look quite similar (see,
for example, Umeda et al. \cite{Umeda94} and Schaab et
al. \cite{Schaab95a}). We shall therefore study only pion
condensation.

Above threshold density for pion condensation $n^{\pi}_c$, the
contribution of eq.~(\ref{GU}) to the neutrino emissivity of the MU
process is to be supplemented by the corresponding pion Urca processes
(cf.  Maxwell et al. \cite{Maxwell77}, \cite{Voskresenskii84}).  For
the latter processes we use a simplified expression including the
nucleon--nucleon correlation effect in the $\pi NN$ vertices
(\cite{Voskresenskii84}, Migdal et al. \cite{Migdal90}),
\begin{eqnarray}\label{cond}
  \epsilon_{\nu}^{(PU)} &\simeq &1.5\times 10^{27}~ \frac{p_{Fn}(n
    )}{m_\pi } \left( \frac{m^{\ast}_N}{m_N} \right)^2 \Gamma^4~
  T^{6}_9 ~\mbox{sin}^2 \theta~ \nonumber \\
  &\times & {\rm\frac{erg}{cm^3~sec}},
\end{eqnarray}
for densities $n>n^{\pi}_c$. Here the neutron Fermi momentum is
expressed in units of $m_\pi$, and $\sin\theta \simeq
\sqrt{2|\tilde{\omega}^2|/m_\pi^2}$ for $\theta \ll 1$, whereas at
large densities ($n \gg n^{\pi}_c$) a well--developed condensate
implies $\theta\rightarrow \pi /2$. We shall use an ansatz that
interpolates between these two extremes. Since $\pi^{+,-}$
condensation probably reduces the energy gaps of the superfluid states
by an order of magnitude (see Takatsuka \& Tamagaki
\cite{Takatsuka80a}), we assume, for the sake of simplicity, that
superfluidity vanishes above $n_c$. Finally we note that though the PU
processes have genuinely one--nucleon phase--space volumes, their
contribution to the resulting emissivity is suppressed relative to the
direct Urca by an additional $\Gamma^2$ vertex factor due to existence
of the additional $\pi NN$ vertex in the former case.

The modifications of energy density and pressure due to the pion
condensation are taken into account as
\begin{equation}\label{encond}
  \rho_\pi \simeq -\frac{\mid \tilde{\omega}^2 (n )\mid \mbox{sin}^2
    (\theta) }{2}\times m_{\pi}^2\,, \quad P_\pi =n^2
  \frac{\partial}{\partial n}\left( \frac{\rho_\pi}{n}\right)\,,
\end{equation} where, again, $\tilde{\omega}^2$ is given in the units
$m_\pi^2$.

%%%%%%%%%% CALCULATION METHOD AND INPUT PHYSICS %%%%%%%%%%%%%%%%% 
\section{Stellar Composition and Structure}\label{sec:input}

The general relativistic equations of stellar structure and thermal
evolution (cf. Thorne \cite{Thorne77}) were solved via a numerical
code based on an implicit finite difference scheme handled by a
Newton-Raphson algorithm. Details can be found in Schaab et
al. (\cite{Schaab95a}).  The physical input quantities, are summarized
in Table \ref{tab:inputs}.  Besides the new neutrino--emission
processes discussed in this paper, we include in the simulations the
traditional neutrino processes too, which are discussed in greater
detail by Schaab et al. (\cite{Schaab95a}).\footnote{The two-nucleon
neutral current processes (bremsstrahlung-processes) are also modified
by in-medium effects (\cite{Voskresenskii86}). Their contribution is
however much smaller than the corresponding contribution of the
MU-VS86 process. These modifications are therefore not implemented
here.}

\begin{table*}
\caption[]{Input quantities used for the cooling
    simulations \label{tab:inputs}}
\begin{tabular}{ll}
\hline\noalign{\smallskip}
Parameter & References \\
\noalign{\smallskip}\hline\noalign{\smallskip}
  Equations of state: \\
  \quad crust        & Baym et al. (\cite{Baym71}), 
                       Negele \& Vautherin (\cite{Negele73}) \\
  \quad core         & see Table \ref{tab:eos} \\
  \hline
  Superfluidity      & see Table \ref{tab:sf} \\
  \hline
  Heat capacity      & Van Riper (\cite{VanRiper91}), 
                       Shapiro \& Teukolsky (\cite{Shapiro83}) \\
  \hline
  Thermal conductivities: \\
  \quad crust        & Itoh et al. (\cite{Itoh84a}), Itoh (\cite{Itoh83a}),
                       Mitake et al. (\cite{Mitake84}) \\
  \quad core         & Gnedin \& Yakovlev (\cite{Gnedin95a}) \\
  \hline
  Neutrino emissivities: \\
  \quad pair-, photon-, plasma-processes & Itoh et al. (\cite{Itoh89}) \\
  \quad bremsstrahlung in the crust      & Itoh \& Kohyama (\cite{Itoh83b}),
                                     Pethick \& Thorsson (\cite{Pethick94b}) \\
  \quad bremsstrahlung in the core       & FM79,
                                           Maxwell (\cite{Maxwell79}) \\
  \quad MU-FM79                  & FM79 \\
  \quad or instead: MU-VS86      & eq. (\ref{GU}), \cite{Voskresenskii86}\\
  \quad NPBF and PPBF processes  & eq. (\ref{NPFB},\ref{PPFB}), Voskresensky \& Senatorov
                                        (\cite{Voskresenskii87a}) \\
  \quad DU-process               & eq. (\ref{eq:DU}), Lattimer et al. \cite{Lattimer91}\\
  \quad PU-process               & eq. (\ref{cond}), \cite{Voskresenskii84}\\
\hline
  photosphere        & Van Riper (\cite{VanRiper88}) \\
\noalign{\smallskip}\hline
\end{tabular}
\end{table*}

%..................... Equation of state ....................................
\subsection{Equation of State}\label{ssec:eos}

\begin{figure}
%\picplace{6.7cm}
\centering\psfig{figure=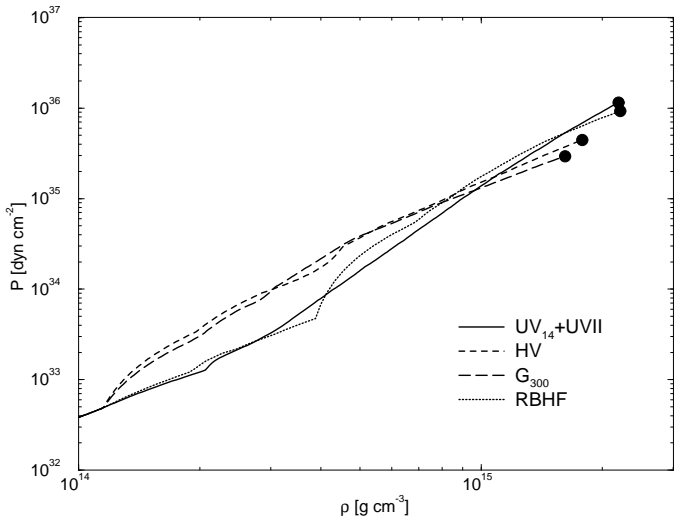,height=8.8cm,angle=-90}
\caption[]{Equations of state (pressure versus energy density)
  used in this paper.  The dots refer to the maximal central densities
  reached in the most massive neutron stars constructed for these
  EOS's.  The modifications caused by a pion condensate alter the EOS's
  only insignificantly on this scale and are thus not shown.
  \label{fig:eos}}
\end{figure}

\begin{figure}
%\picplace{6.7cm}
\centering\psfig{figure=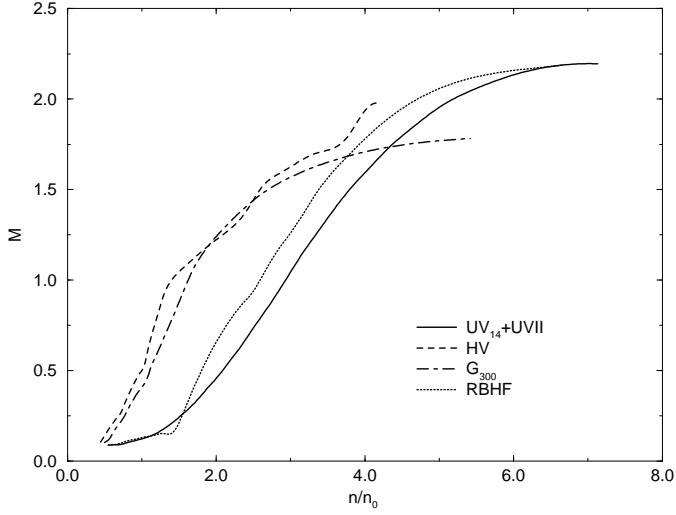,height=8.8cm,angle=-90}
\caption[]{Star masses in units of solar mass versus central 
  star density for the EOS's used in this paper.
\label{fig:mass}}
\end{figure}

For the outer and inner crust we adopt the equations of state of Baym
et al.  (\cite{Baym71}) and Negele \& Vautherin (\cite{Negele73}).
The transition density between the ionic crust and the core of a
neutron star is assumed to be $\rho_{\rm{tr}}=1.7\times 10^{14}\gccm$
(Pethick et al. \cite{Pethick95}). The high density matter in the
cores of neutron stars is described by a collection of modern
equations of state, which consists of a non-relativistic
Schr\"odinger-based model as well as three relativistic field
theoretical ones. The details of these models are summarized in Table
\ref{tab:eos}. One of the significant differences between these
equations of state is that the non-relativistic model treats neutron
star matter as being composed of neutrons and protons only (which are
in $\beta$--equilibrium with leptons).  The relativistic models
account for all hyperon states that become populated in the cores of
neutron stars as predicted by theory (Glendenning
\cite{Glendenning85a}). Dynamical two-nucleon correlations calculated
from the relativistic scattering T-matrix in matter are contained in
the relativistic Brueckner-Hartree-Fock (RBHF) equation of state. The
underlying one-boson-exchange interaction is Brockmann's potential
``B''.  At densities larger than about three times normal nuclear
matter density it has been joined with a relativistic Hartree-Fock
(RHF) equation of state (Huber et al. \cite{Huber94,Huber96a}). The
non-relativistic treatment of the EOS leads generally to a softer EOS
than the relativistic treatment. However, since in the
non-relativistic case one takes only nucleons and leptons into account
the relativistic EOS's with hyperons become almost as soft as the
non-relativistic EOS's. Both treatments therefore mainly differ in the
resulting composition. This can be important for the nucleon and
hyperon direct Urca (see Sect.
\ref{sec:du}).  The equations of state are shown in Fig. \ref{fig:eos}
and the corresponding neutron star sequences in Fig. \ref{fig:mass}.

\begin{table*}
\caption[]{Collection of nuclear equations of state (EOSs).
           Abbreviations: nvar=nonrelativistic
             variational, RH=relativistic Hartree, RHF=relativistic
             Hartree-Fock, RBHF=relativistic Brueckner-Hartree-Fock.
\label{tab:eos}}
\begin{tabular}{ccccc}
\hline\noalign{\smallskip}
EOS & Composition & Interaction & Method & Ref. \\
\noalign{\smallskip}\hline\noalign{\smallskip}
UV$_{14}$+UVII &p,n,e,$\mu$ &Urbana V$_{14}$+VII  &nvar
    & Wiringa et al. (\cite{Wiringa88}) \\
HV &p,n,$\Lambda,\Sigma,\Xi$,e,$\mu$   &$\sigma,\omega,\rho$ 
   &RH   & Weber \& Weigel (\cite{Weber89}) \\
RBHF(B)+HFV  &p,n,$\Lambda,\Sigma,\Xi,\Delta$,e,$\mu$  
&$\sigma,\omega,\pi,\rho,\eta,\delta$  &RBHF, RHF  & Huber et al. (\cite{Huber94}) \\
G$_{300}$  &p,n,$\Lambda,\Sigma,\Xi$,e,$\mu$   &$\sigma,\omega,\rho$
           &RH   & Glendenning (\cite{Glendenning89}) \\
\noalign{\smallskip}\hline
\end{tabular}
\end{table*}

%.................... Superfluidity .......................................
\subsection{Superfluidity}

\begin{figure}
%\picplace{6.7cm}
\centering\psfig{figure=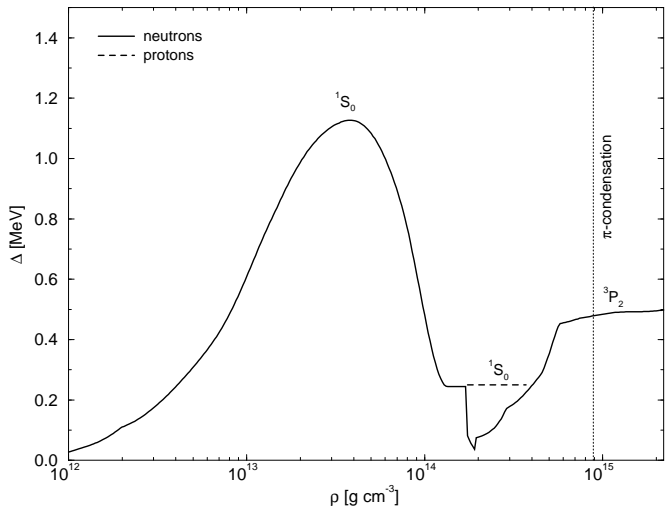,height=8.8cm,angle=-90}
\caption[p0995b_sfgap.ps]{Superfluid energy gaps computed for the HV EOS.
We neglect superfluidity in the models where pion condensation is allowed 
for densities $n>n^{\pi}_{\rm c}=3n_0$ (dotted line).
\label{fig:sf}}
\end{figure}

The following superfluid regions inside neutron stars are taken into account:
\begin{itemize}
\item $7\times 10^{11}\gccm \le \rho \le 2\times 10^{14}\gccm$: neutrons in
  \sfs-pair state,
\item $2\times 10^{14}\gccm \le \rho \le 4\times 10^{14}\gccm$: protons in
  \sfs-pair state, and
\item $2\times 10^{14}\gccm \le \rho \le 5\times 10^{15}\gccm$: neutrons in
  \sfp-pair state.
\end{itemize} 
The proton gap is taken from Wambach et al. (\cite{Wambach91a}), those
of \sfs-- and \sfp--paired superfluid neutrons are from Ainsworth et
al. (\cite{Ainsworth89a}) and Amundsen \& {\O}stgaard
(\cite{Amundsen85}), respectively. Details are given in table
\ref{tab:sf}. It should be emphasized that the extension of the
superfluid density regime depends on the equation of state.  The above
values are computed for the HV equation of state, and
Fig.~\ref{fig:sf} displays the profiles of the superfluid gaps for
this particular model.

\begin{table*}
  \caption[]{Maximal gap energies, $\Delta_{\rm sf}^{\rm max}$,
    critical temperatures, $T_c^{\rm max}$, and density ranges,
    $\rho$, in neutron star matter computed for the HV equation of
    state \label{tab:sf}}
\begin{tabular}{cccc} 
\hline\noalign{\smallskip}
& Neutron \sfs & Neutron \sfp & Proton \sfs \\
\noalign{\smallskip}\hline\noalign{\smallskip}
$\Delta^{\rm{max}}_{\rm{sf}}$ [MeV]  &1.13      &0.62      &0.25 \\
$T^{\rm{max}}_c$         [$10^9$~K]  &7.4       &0.8       &1.6  \\
$\rho$                    [$\gccm$]  &$7\times 10^{11}-2\times10^{14}$   
& $2\times 10^{14}-5\times 10^{15}$  & $2\times10^{14}-4\times10^{14}$ \\  
Reference   & Ainsworth et al. (\cite{Ainsworth89a})  
& Amundsen \& {\O}stgaard (\cite{Amundsen85})  
& Wambach et al. (\cite{Wambach91a}) \\
\noalign{\smallskip}\hline
\end{tabular}
\end{table*}

Because the pairing gaps are not too accurately known, we shall also
study cooling scenarios of superfluid neutron stars where the
magnitudes of the gaps are varied in an {\it ad hoc} fashion about
their theoretically determined values. This procedure may serve to
reveal the dependence of our theoretical cooling scenarios on the
uncertainties associated with the superfluid gaps.

As already mentioned in Sect. \ref{sec:emis.smu}, the superfluid
gaps are reduced by an order of magnitude in $\pi^{+,-}$ condensed
matter (see Takatsuka \& Tamagaki \cite{Takatsuka80a}). We therefore
neglect superfluidity in pion condensed matter completely. In passing
we emphasize that in a more elaborate treatment, one needs to pay
attention to the effects of interactions between $\pi^{+,-}$ and
$\pi^{0}$ condensates, simultaneous appearance of $K^{-}$ and
$\bar{K}^{0}$ condensates (Brown et al. \cite{Brown94}, Kolomeitsev et
al. \cite{Kolomeitsev95}) which affect the neutron and proton gaps
differently. Furthermore, neutron--proton pairing in the
$^3D_2$--triplet state for almost symmetric nuclear matter\footnote{In
neutron star matter with hyperons based on relativistic EOS's the
neutron and proton fractions tend to each over with increasing density
and are almost equal in the high density limit (cf. Weber \& Weigel
\cite{Weber89} and Glendenning \cite{Glendenning89}). One must 
distinguish this case from the nonrelativistic treatment without
hyperons, where in the presence of kaon condensation isospin symmetric
matter is preferred (cf. Thorsson et al. \cite{Thorsson94a}).}  may
dominate the traditionally considered $^3P_2$ pairing (Alm et
al. \cite{alm96}).

Finally when performing numerical cooling calculations for superfluid
neutron stars, we follow the traditional procedure by suppressing the
two--nucleon reaction neutrino emissivities, the thermal conductivity,
and the heat capacity of the nucleonic constituents of the stellar
matter by factors which behave like $\exp(-\Delta_{\rm{n\, (p)}}
/k_{\rm{B}}T)$ for $T\ll T_c$, where $\Delta_{\rm{n\, (p)}}$
denotes the gap, $T$ the temperature (see Maxwell
\cite{Maxwell79}, Gnedin \& Yakovlev \cite{Gnedin95a})\footnote{
  For the \sfp~state this is rigorously true
  only for isotropic system (see Anderson \& Morel \cite{Anderson61},
  Muzikar \cite{Muzikar80}, and Page \cite{Page95}).}.

%%%%%%%%%%%%%%%%% Results and Observations %%%%%%%%%%%%%%%%%%%%%%%%%%%%%%%%%%
\section{Results and Discussion}

%................ Observational data ........................................
\subsection{Observational Data} \label{subsec:observations}

Compact X-ray sources have been detected by the X-ray observatories
Einstein, EXOSAT and ROSAT during the last two decades. Among them, 14
X--ray sources -- observed at least by one of these satellites -- were
identified as radio--pulsars (with the exception of Geminga, which is
known to be radio--quiet).  The information obtained from these
detections is not always sufficient to extract the effective surface
temperature of the corresponding pulsar. Therefore this sample of
pulsars is divided into three different categories (see {\"O}gelman
\cite{Oegelman95a}):
\begin{enumerate}
\item The detection of three pulsars (PSR's 1706-44, 1823-13, and
  2334+61) contain too few photons for spectral fits. The specified
  luminosities are calculated by using the totally detected photon
  flux.  These pulsars are marked with triangles in the figures that
  will be discussed below.
\item The spectra of seven pulsars, including the Crab pulsar (PSR
  0531+21), can only be fitted by a power--law--type spectrum, or by a
  blackbody spectrum with very high effective temperature and
  effective areas much smaller than the neutron star surface. Their
  X-ray emission is predominated by magnetospheric emission.
  Therefore, the temperatures, determined from the spectral fits, are
  probably too high. Pulsars of this type are marked with dots.
\item Finally, there are four pulsars, i.e., 0833-45 (Vela), 0656+14,
  0630+18 (Ge\-min\-ga), and 1055-52, allowing two--component spectral
  fits. The softer blackbody component is believed to correspond to
  the actual surface emission of the neutron star, while the harder
  blackbody (or power--law) component may be due to magnetospheric
  emission.  These pulsars are marked with squares.
\end{enumerate} 
An overview of the observed luminosities and pulsar ages is contained
in Table \ref{tab:observations}.

\begin{table*}
  \caption[]{Luminosities, $L$, and spin-down ages, $\tau$, of pulsars}
  \label{tab:observations}
\begin{tabular}{ccccc}
\hline\noalign{\smallskip}
Pulsar & Name & $\log\tau$ [yr] & $\log L$ [erg/s] & Reference \\
\noalign{\smallskip}\hline\noalign{\smallskip}
  \multicolumn{5}{c}{\small not enough data available for spectral analysis} \\
  1706-44 & & 4.25 & $32.8\pm 0.7$ & Becker et al. (\cite{Becker92a}) \\
  1823-13 & & 4.50 & $33.2\pm 0.6$ & Finley \& {\"O}gelman (\cite{Finley93b})\\
  2334+61 & & 4.61 & $33.1\pm 0.4$ & Becker (\cite{Becker93b}) \\
  \hline
  \multicolumn{5}{c}{\small power-law-type spectra or spectra with only
a high temperature component} \\
  0531+21 & Crab & 3.09 & $33.9\pm 0.2$ & Becker \& Aschenbach (\cite{Becker95a})\\
  1509-58 & SNR MSH 15-52 & 3.19 & $33.6\pm 0.4$  
  & Seward et al. (\cite{Seward83a}), Trussoni et al. (\cite{Trussoni90a}) \\
  0540-69 & & 3.22 & $36.2\pm 0.2$ & Finley et al. (\cite{Finley93a}) \\
  1951+32 & SNR CTB 80 & 5.02 & $33.8\pm 0.5$ & Safi-Harb \& {\"O}gelman (\cite{SafiHarb95a}) \\
  1929+10 & & 6.49 & $28.9\pm 0.5$ & Yancopoulos et al. (\cite{Yancopoulos93}),
  {\"O}gelman (\cite{Oegelman95a}) \\
  0950+08 & & 7.24 & $29.6\pm 1.0$ & Seward \& Wang (\cite{Seward88a}) \\
  J0437-47 & & 8.88 & $30.6\pm 0.4$ & Becker \& Tr{\"u}mper (\cite{Becker93c}) \\
  \hline
  \multicolumn{5}{c}{\small spectrum dominated by a soft component} \\
  0833-45 & Vela & 4.05 & $32.9\pm 0.2$ & {\"O}gelman et al. (\cite{Oegelman93a}) \\
  0656+14 & & 5.04 & $32.6\pm 0.3$ & Finley et al. (\cite{Finley92a}) \\
  0630+18 & Geminga & 5.51 & $31.8\pm 0.4$ & Halpern \& Ruderman (\cite{Halpern93a}) \\
  1055-52 & & 5.73 & $33.0\pm 0.6$ & {\"O}gelman \& Finley (\cite{Oegelman93b}) \\
\noalign{\smallskip}\hline
\end{tabular}
\end{table*}

%............... Impact of softening ........................................
\subsection{Impact of Softening of Pion Exchange Modes and Intermediate
  Pion Decay on Cooling}

\begin{figure}
%\picplace{6.6cm}
\centering\psfig{figure=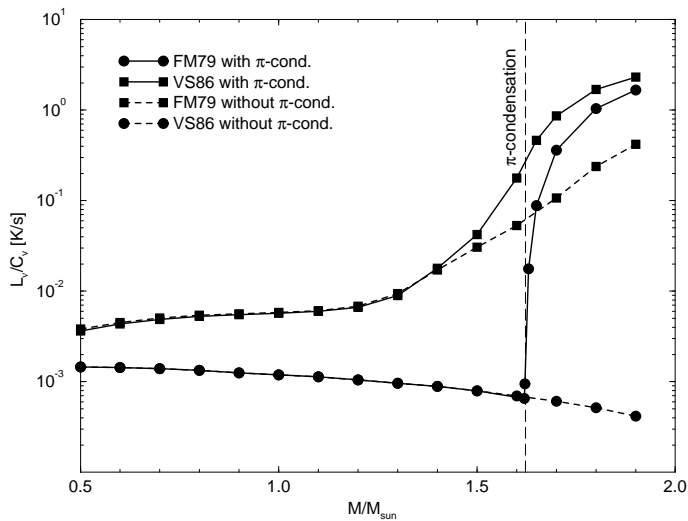,height=8.8cm,angle=-90}
\caption[]{Cooling rate due to neutrino emission as a function of 
  star mass for a representative temperature of $T=3\times
  10^8$~K. The solid curves refer to cooling via MU-FM78+PU and
  MU-VS86+PU in pion--condensed (threshold density $n_c=3n_0$)
  matter. The dashed curves refer to cooling without pion
  condensation. Superfluidity is neglected here.  \label{fig:rate}}
\end{figure}

\begin{figure*}
%\picplace{10.4cm}
\centering\psfig{figure=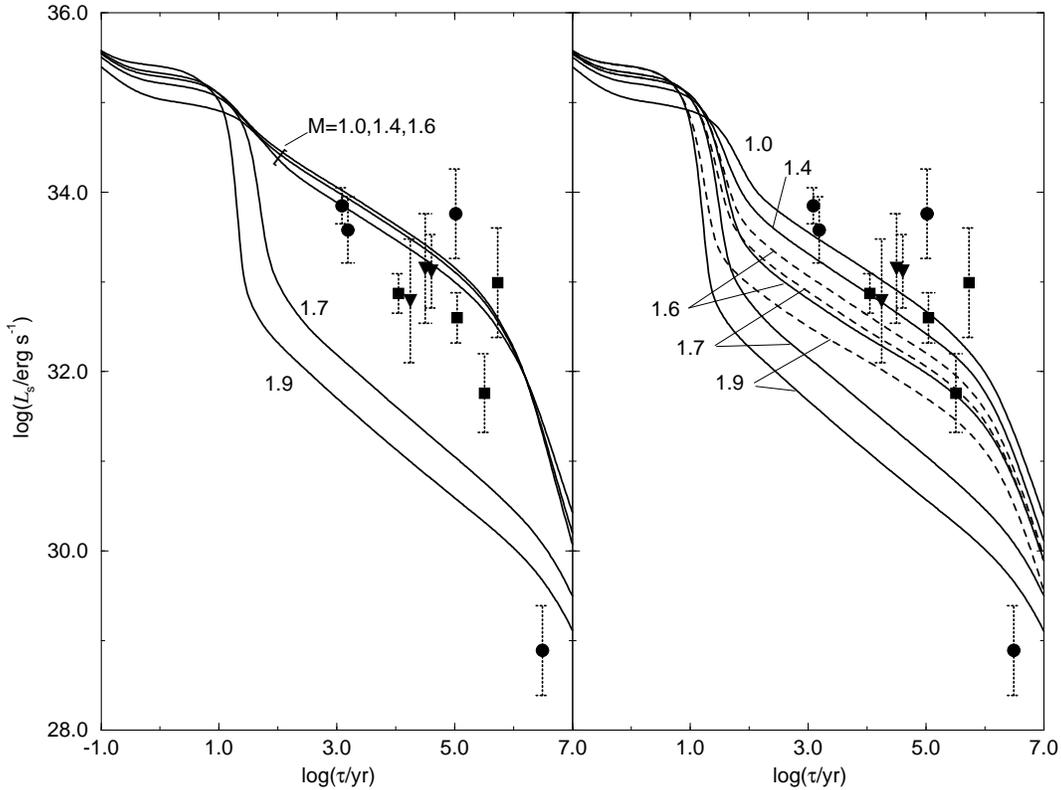,height=14cm,angle=-90}
\caption[]{Cooling of non--superfluid neutron star models of 
  different masses constructed for the HV EOS. The two graphs refer to
  cooling via MU-FM79+PU (left) and MU-VS86+PU (right). In both cases,
  pion condensation is taken into account for the solid curves where
  $n>n^{\pi}_c=3n_0$. The dashed curves in the right graph refer to
  the $\tilde{\omega}^2$ parametrization without pion condensation. The
  observed luminosities are labeled in
  Fig. \ref{fig:rbhf_all_sf}. \label{fig:hvh_nsf}}
\end{figure*}

\begin{figure*}
%\picplace{10.4cm}
\centering\psfig{figure=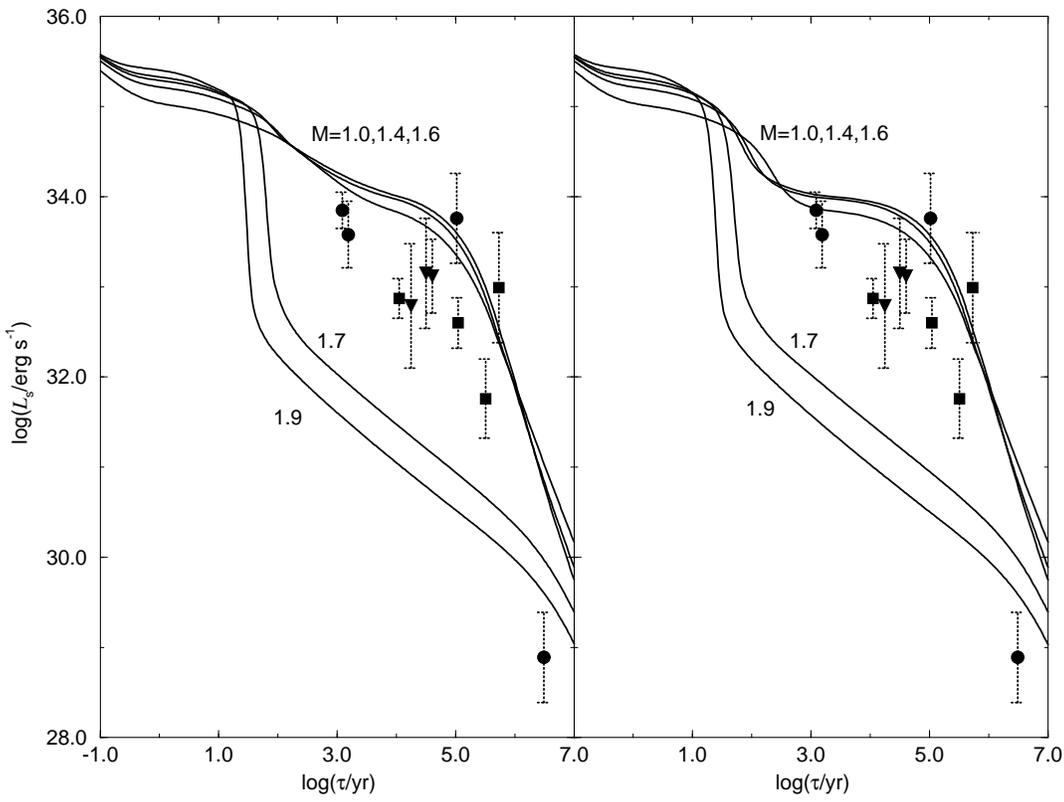,height=14cm,angle=-90}
\caption[]{Same as Fig. \ref{fig:hvh_nsf} but for superfluid 
  neutron star models. NPBF and PPBF are not incorporated. The
  observed data are labeled in
  Fig. \ref{fig:rbhf_all_sf}. \label{fig:hvh_sf}}
\end{figure*}

In Figs. \ref{fig:rate}--\ref{fig:hvh_sf} we show the impact of the
medium effects on the cooling of neutron stars of different
masses. Fig. \ref{fig:rate} compares the mass dependence of the
neutrino cooling rates $L_\nu /C_V$ associated with MU-FM79 and
MU-VS86 for non--superfluid matter. For the solid curves, the neutrino
emissivity in pion--condensed matter (threshold density
$n^{\pi}_c=3n_0$) is taken into account according to eq. (\ref{cond})
and (\ref{GU}) with the parameter $\alpha =\sqrt{2}$. The dashed
curves correspond to the case where no pion condensation is
allowed. As one sees the medium polarization effects included in
MU-VS86 may result in three order of magnitude increase of the cooling
rate for the most massive stars. Even for stars of a low mass the
cooling rate of MU-VS86 is still few times larger than for MU-FM79
because even in this case the more efficient rate is given by the
reactions shown by the right diagram in Fig.1.  The cooling rates of
the $1.8M_\odot$ mass models with and without pion condensate differ
only by a factor of 5.

One sees that the cooling rates for the FM79
result vary significantly (by about three orders of
magnitude!) only over a rather small mass range from the models
without pion condensation to the ones with pion condensation.  The
cooling rates are rather independent of star mass below the critical
mass value (i.e., $M=1.63\,M_\odot$) at which the transition into the
pion--condensed phase occurs, and show a flattening behavior
above. The resulting cooling behavior of neutron stars of several
selected masses is shown in the left graph of
Fig. \ref{fig:hvh_nsf}. These curves are quite similar for models
without pion condensation.  Stars which are sufficiently heavy
such that a pion condensate can develop in their cores evolve along
cooling tracks that are rather different from the former, provided the
stars are older than about 10 years.  We illustrate this for two
representative star masses, $M=1.7$ and $M=1.9\,M_\odot$. Depending on
star mass, the resulting photon luminosities are basically either too
high or too low to account for the bulk of observed pulsar
luminosities, which tends to be a general feature of cooling
calculations no matter what kind of enhanced cooling mechanism are
being studied (cf. Schaab et al. \cite{Schaab95a}).

The situation changes if the modified Urca process with the medium
modifications of the pion-exchange interaction, appropriate vertex
corrections, and the radiation from the intermediate states (MU-VS86) is
included. Now the cooling rates vary smoothly with star mass (see Fig.
\ref{fig:rate}) such that the gap between standard and enhanced
cooling is washed out. More quantitatively, by means of varying the
star mass between 1.0 and 1.6 $M_\odot$, one achieves agreement with a
large number of observed data points, see right graph of
Fig. \ref{fig:hvh_nsf}. This is true for both choices of the
$\tilde{\omega}^2$ parametrization, independently whether a pion
condensation can occur or not. The two parametrizations differ only
in the range which is covered by the cooling curves. The only pulsars
which do not agree with the cooling curves are the hot PSR's 1055-52
and 1951+32 and the rather cold object PSR 1929+10. The high
luminosities of the former two may be due to internal heating
processes, which leads to delayed cooling for star ages $t>10^5$~yr
(see for example Shibazaki \& Lamb
\cite{Shibazaki89}, Sedrakian \& Sedrakian \cite{Sedrakian93}, 
Reisenegger \cite{Reisenegger95a}, Van Riper et
al. \cite{VanRiper95a}, Schaab et al. \cite{Schaab96c}).  To achieve
agreement with the extremely low--luminosity object PSR 1929+10, the
inclusion of strong magnetic fields in the atmosphere (see Van Riper
\cite{VanRiper91}) or 
other fast cooling processes, like direct Urca etc. seems necessary
(cf. Fig. \ref{fig:hvh_all_sf} and Schaab et al. \cite{Schaab95a}).

Next we shall study the modified Urca process in superfluid
matter. Results are displayed in Fig. \ref{fig:hvh_sf}.  The processes
NPBF and PPBF, which will be discussed later, have been artificially
forbidden. The value of the neutron \sfp~gap from Amundsen \&
{\O}stgaard (\cite{Amundsen85}) is rather large such that the
difference between slow (MU-FM79, left graph) and intermediate
(MU-VS86, right graph) cooling is almost indistinguishable.
Superfluidity strongly suppresses the neutrino emission rates from
stars containing a superfluid core, which thus delays cooling. As it
was the case for standard cooling, the thermal evolution of superfluid
stars is rather insensitive against variations of the star's mass.
However, as soon as charged pions condense in the core of a neutron
star, superfluidity disappears and the much higher neutrino emission
rates cause the cooling curves to drop down by more than two orders of
magnitude. For the model which allows for pion condensation, this
occurs when the star's mass is varied from 1.6 to 1.7 $M_\odot$.  In
principle, it appears possible to achieve agreement between the body
of observed cooling data and the cooling simulations by properly
varying the star's mass in an extremely narrow range. Surely, such a
procedure is not very satisfying because it requires an extreme
fine--tuning of the mass of a star.  As we shall see below, the
superfluid state allows for other neutrino--emitting processes which
decisively influence the cooling of stars and thus provide other, less
stringent conditions by means of which agreement with the observed
data can be achieved.

%.................... Superfluid pair breaking ..............................
\subsection{Superfluid Pair Breaking and Formation}

\begin{figure}
%\picplace{10.5cm}
\centering\psfig{figure=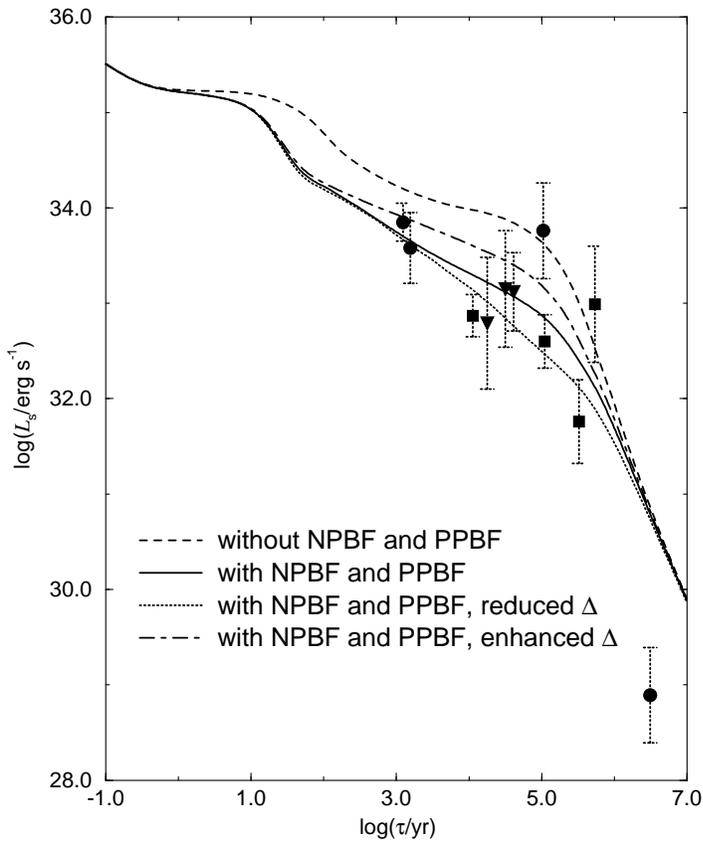,width=8.8cm}
\caption[]{Cooling of neutron stars of mass 
  $M=1.4M_\odot$ constructed for the HV EOS.  The curves correspond to
  standard cooling (MU-VS86 and PU ignored) with (solid curve) and without
  (dashed curve) NPBF+PPBF.  The gap energy of the \sfs --neutron--pairing is arbitrarily
  enhanced (dotted curve) or reduced (dashed-dotted curve),
  respectively by a factor of 2 in order to demonstrate the dependency of the rates on the gap
  energy. The observed data are labeled in Fig.
  \ref{fig:rbhf_all_sf}.  \label{fig:hvh_sfpair}}
\end{figure}

To demonstrate the relevance of the superfluid pair breaking and
formation processes for the cooling of neutron stars, in the first
step, we implement these two processes into standard cooling
calculations. In these calculations the FM79 result for the modified
Urca process is used. No further processes, like DU or PU is taken
into account here.  The results are shown in
Fig. \ref{fig:hvh_sfpair}.  One sees that NPBF and PPBF reduce the
photon luminosity of neutron stars significantly for star ages between
about $1 <t <10^6~\rm{yr}$.  The acceleration of the cooling is mostly
due to the superfluid transition of the neutrons in the crust since
the gap energy for this transition is 1.13~MeV compared to 0.25~MeV
for the protons in the core (see Tab. \ref{tab:sf}) and the value of
$\Delta/T$ at $T=T_{\rm c}$ is 0.57 compared to 0.12 for the neutrons
in the core (see Sec. \ref{ssec:pair}). One can also see that the
importance of the NPBF process increases with decreasing gap energy of
the \sfs --neutron--pairing. For $t>10^{6}~\rm{yr}$ the influence of
NPBF and PPBF becomes rather weak and cooling via photon emission from
the star's surface becomes the dominant process. As a result, the
solid and dashed curves in Fig. \ref{fig:hvh_sfpair} are almost
identical for $t>10^{6}~\rm{yr}$.

%.................... Final Results ..............................
\subsection{Final Results and Influence of Equation of State}
\label{sec:finres}

\begin{figure*}
%\picplace{10.4cm}
\centering\psfig{figure=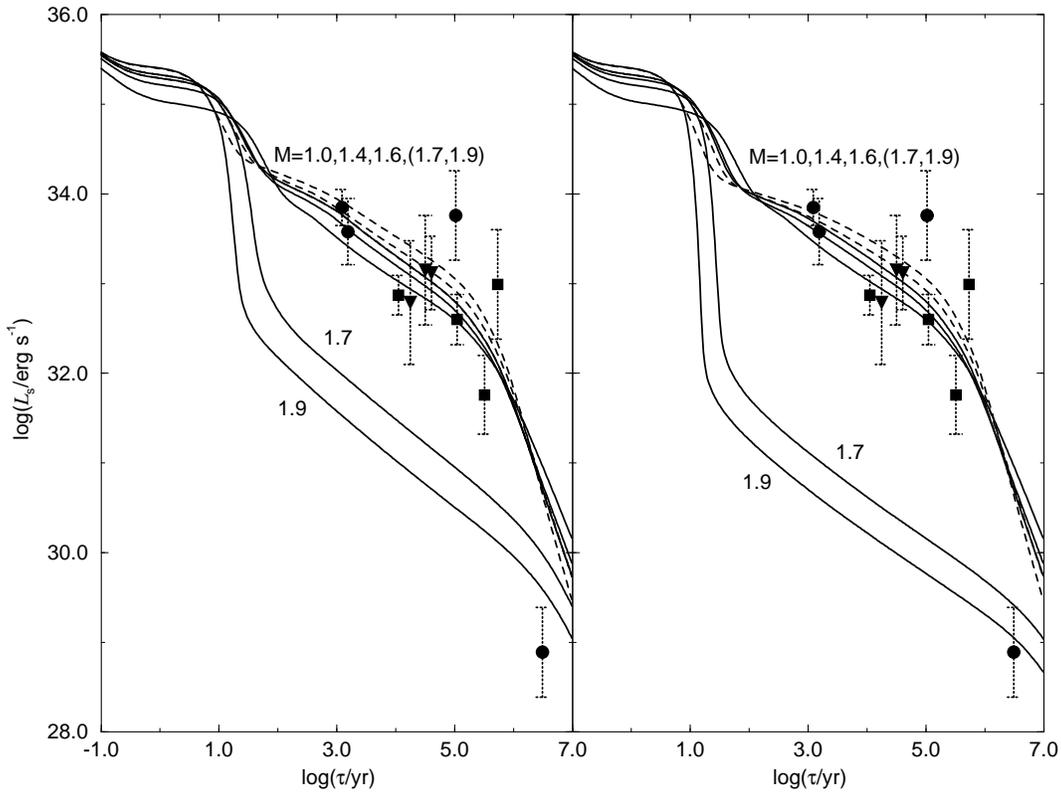,height=14cm,angle=-90}
\caption[]{Cooling of neutron stars with different masses 
  constructed for the HV EOS. The cooling processes are MU-VS86, PU (only solid curves), NPBF,
  PPBF, and DU (only in the right graph). The dashed curves refer to the $M=1.7$ and $1.9M_{\odot}$ models without pion condensate. The observed data are labeled in Fig.
  \ref{fig:rbhf_all_sf}. \label{fig:hvh_all_sf}}
\end{figure*}

We turn now to cooling simulations where the MU, NPBF, PPBF, DU and PU
take place simultaneously.  Fig. \ref{fig:hvh_all_sf} shows the
cooling tracks of stars of different masses, computed for the
HV equation of state. The DU process is taken into account in the
right graph, whereas it is neglected in the left graph. The solid
curves refer again to the $\tilde{\omega}^2$--parameterization with
phase transition to a pion condensate, the dashed curves to the one
without phase transition. For masses in the range between 1.0 and 1.6
$M_\odot$, the cooling curves pass through most of the data points.
We again recognize a photon luminosity drop by more than two orders of
magnitude of the 1.7 $M_\odot$ mass star with pion condensate, due to
vanishing of the superfluidity as a consequence of pion
condensation. This drop is even larger if the DU is taken into account
(right graph, this allows to account for the photon luminosity of PSR
1929+10).

A comparison with the observed luminosities shows that one
gets quite good agreement between theory and observation if one assumes
that the masses of some of the underlying pulsars are different from
the canonical value, $M=1.4\,M_\odot$. Since the masses of these
pulsars are not known, no further conclusions about the actually
operating cooling mechanism of these pulsars can be drawn yet.  Future
mass determinations of these objects will change the situation.  The
only two stars whose photon luminosities cannot be accounted for by
the new processes studied here are the rather hot pulsars PSR 1055-52
and PSR 1951+32. Again, as mentioned above, their temperatures may be
explained by internal heating.

\begin{figure*}
%\picplace{10.5cm}
\centering\psfig{figure=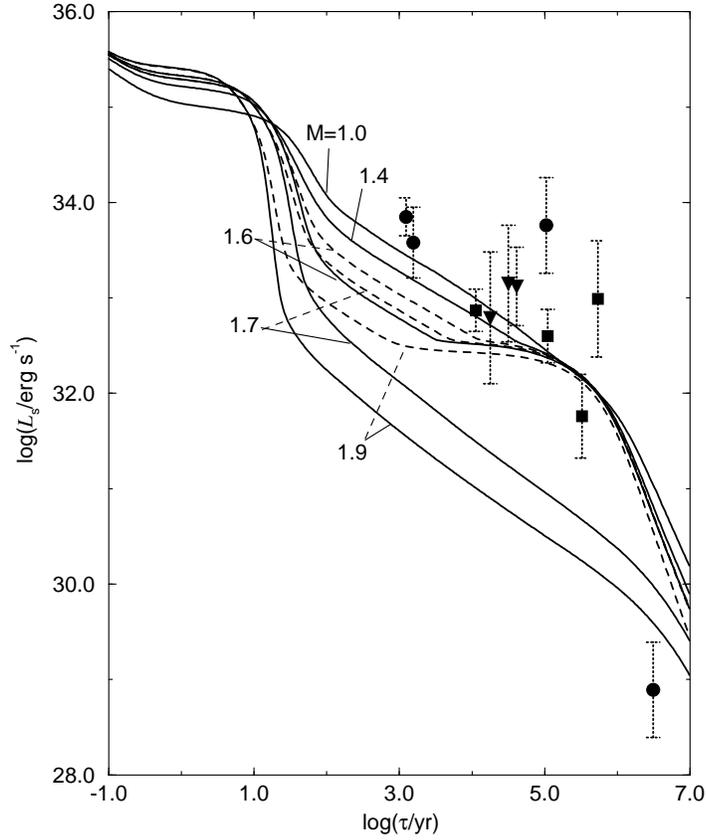,width=8.8cm}
\caption[]{Same as left graph of Fig. \ref{fig:hvh_all_sf} but with the 
  original \sfp ~gap by Amundsen \& {\O}stgaard (\cite{Amundsen85})
  reduced by a factor of 6. The observed data are labeled in Fig.
  \ref{fig:rbhf_all_sf}. \label{fig:hvh_all_sf2}}
\end{figure*}

To explore the impact of somewhat smaller superfluidity gaps on
cooling, we arbitrarily reduced the gap by a factor of 6. Its major
effect is a drop in the cooling curves of the 1.0 to 1.6 $M_\odot$
mass stars and of the 1.7 and 1.9 $M_\odot$ mass stars without pion
condensate (dashed curves). This would lead to better agreement with
the low--temperature pulsar 0630+18 (Geminga) (see
Fig. \ref{fig:hvh_all_sf2}).

\begin{figure}
%\picplace{10.5cm}
\centering\psfig{figure=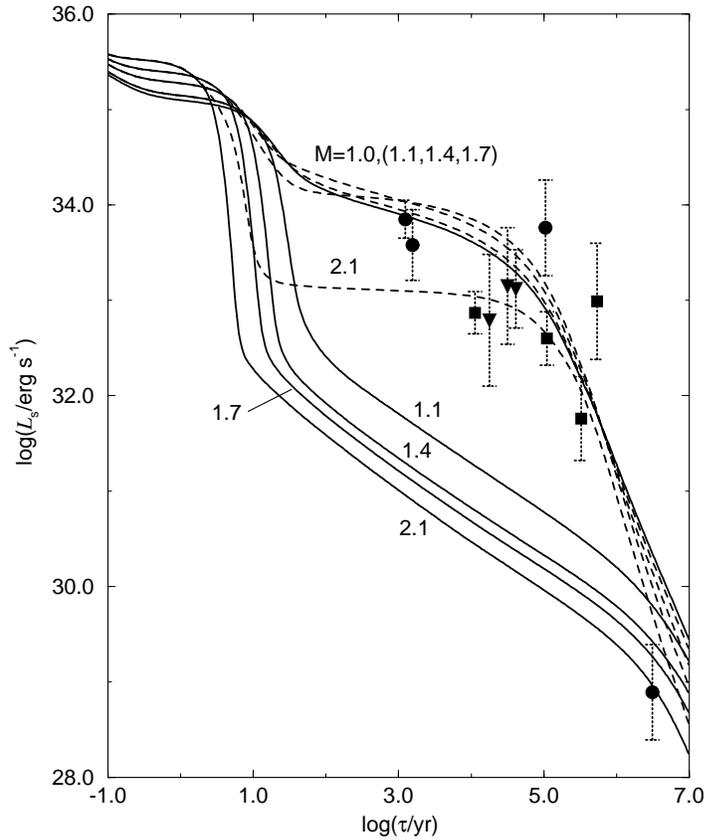,width=8.8cm}
\caption[]{Cooling of neutron stars of different masses 
  constructed for the UV$_{14}$+UVII EOS. The cooling processes are
  MU-VS86, PU (only solid curves), NPBF, and PPBF. The observed data are
  labeled in Fig.  \ref{fig:rbhf_all_sf}. \label{fig:uvu_all_sf}}
\end{figure}

\begin{figure}
%\picplace{10.5cm}
\centering\psfig{figure=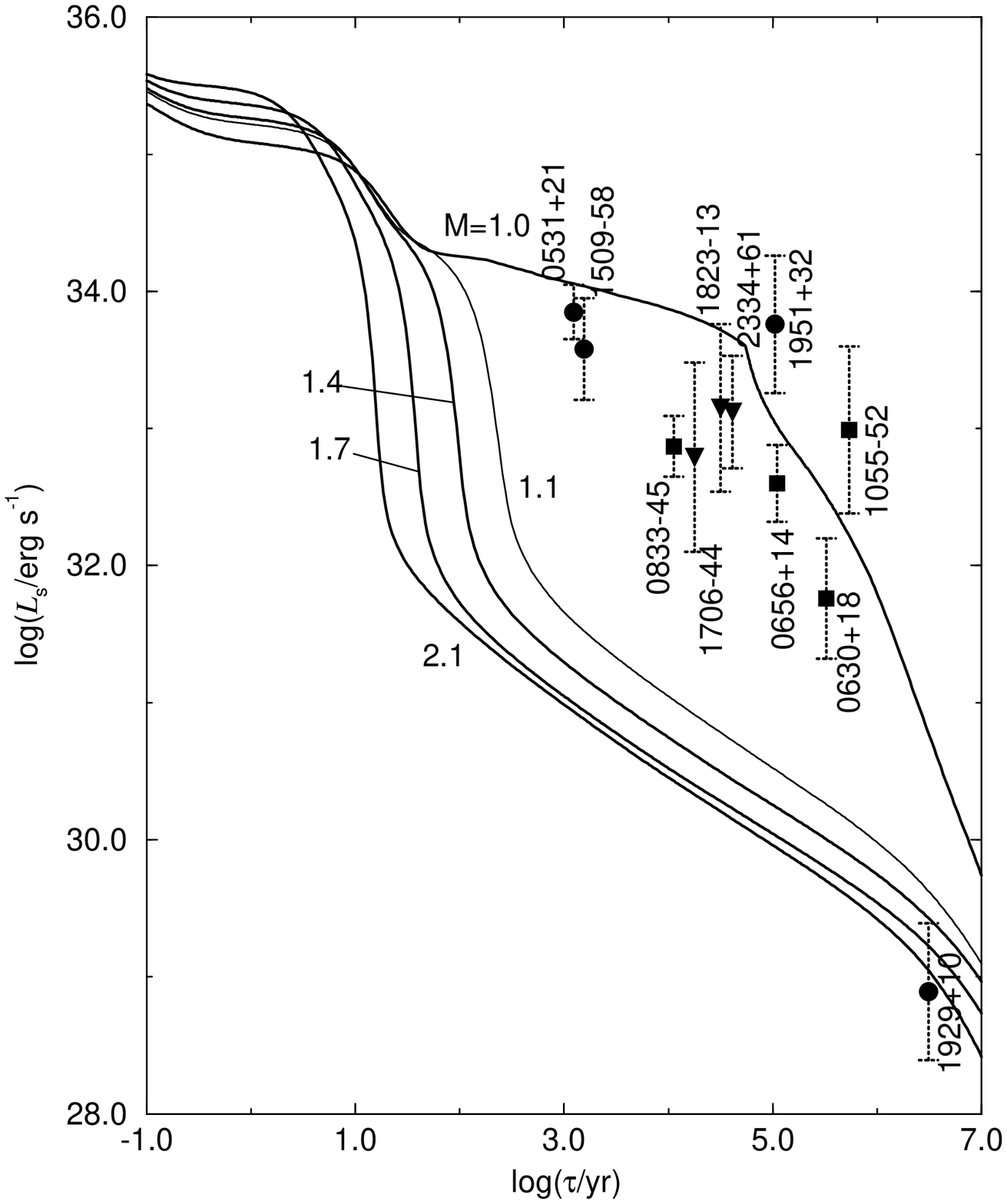,width=8.8cm}
\caption[]{Cooling of neutron stars of different masses 
  constructed for the RBHF EOS. The cooling processes are
  MU-VS86, PU, NPBF, and PPBF. The observed data are
  labeled in Fig.  \ref{fig:rbhf_all_sf}. \label{fig:rbhf_all_sf}}
\end{figure}

\begin{figure}
%\picplace{10.5cm}
\centering\psfig{figure=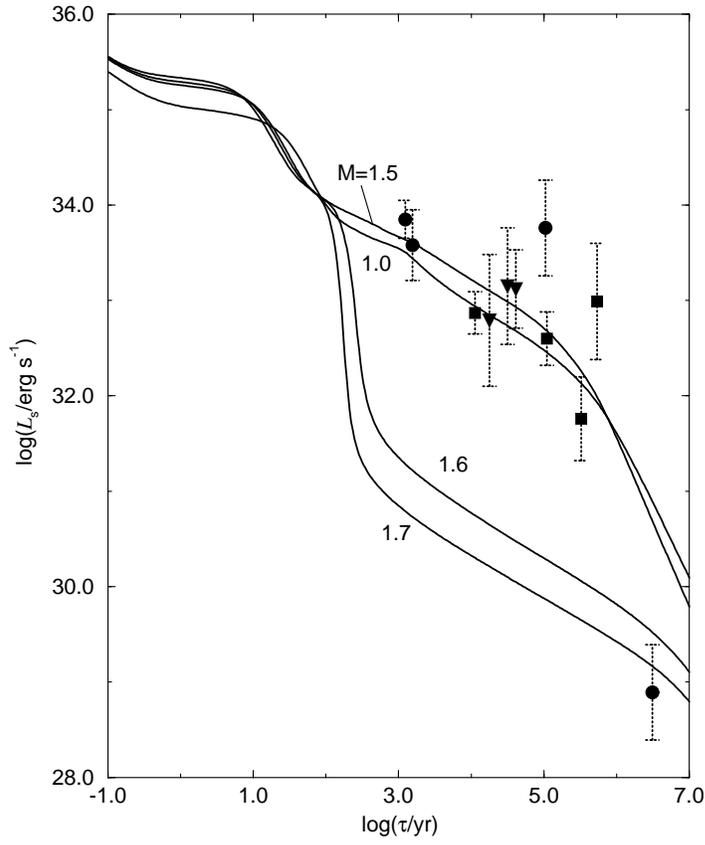,width=8.8cm}
\caption[]{Same as Fig. \ref{fig:rbhf_all_sf} but for the 
  G$^{300}$ EOS.  The observed data are labeled in
  Fig. \ref{fig:rbhf_all_sf}.  \label{fig:g300_all_sf}}
\end{figure}

So far we have selected a particular model for the equation of state
of neutron star matter, i.e. HV, and studied the dependence of the
cooling curves on the star's mass. The influence of different models
for the equation of state on cooling is shown in
Figs. \ref{fig:uvu_all_sf}--\ref{fig:g300_all_sf}. Fig.
\ref{fig:uvu_all_sf} displays the cooling behavior computed for 
a representative non--relativistic model for the equation of state,
UV$_{14}$+UVII. The general qualitative features obtained for HV carry
over to this case, except that the temperature drop occurs now at a
lower mass threshold. For that reason the 1.1 $M_\odot$ mass star is
already anomalously cold. Because of the relatively high neutron
fraction of the UV$_{14}$+UVII compared to relativistic EOS's the
superfluid phase does not reach to the center of the $2.1M_\odot$ mass
star in this particular model of Amundsen \& {\O}stgaard
(\cite{Amundsen85}). This causes the luminosity drop to occur also for
the model without pion condensate (dashed curve). The high neutron
fraction, or identically the low proton fraction, has also the effect
that the DU process cannot occur. 

The cooling curves constructed for RBHF, shown in
Fig. \ref{fig:rbhf_all_sf}, are rather similar to those of
UV$_{14}$+UVII though the underlying microscopic theories employed to
determine these two models for the equation of state are completely
different. We recall that the former, RBHF, is based on a
parameter--free relativistic Brueckner-Hartree-Fock calculation while
the latter, UV$_{14}$+UVII, was derived in the framework of a
non-relativistic many-body variational calculation.  In both cases,
however, dynamical nucleon-nucleon correlations are taken into
account, which -- as a general feature -- tend to soften the equation
of state at intermediate nuclear densities, both in the relativistic
as well as in the non-relativistic case (Fig.
\ref{fig:eos}). For that reason both of these models for the 
equation of state lead to similar slopes of star mass as a function of
central star density (see Fig. \ref{fig:mass}) and similar cooling
behaviors. With respect to the differences between RBHF and
UV$_{14}$+UVII, the former is harder than the latter. Therefore the
magnitude of the temperature drop for increasing star masses is
smaller.  For both of these models, pion condensation set in at rather
low neutron star masses, causing already the rather light neutron
stars ($M\sim 1.1\,M_\odot$) to cool very rapidly. In contrast to the
UV$_{14}$+UVII EOS, the RBHF has large proton fractions at high
densities causing the superfluid phase to reach to the center of the
star for all star masses and allowing for the DU process. The cooling
tracks for the $\tilde{\omega}^2$--parameterization without pion
condensation and the ones with the DU process are similar to the
cooling tracks for the HV EOS. Thus, we can restrict ourselves to the
pion condensation case without DU process.

Besides HV, we employ a second model, labeled G$_{300}$, for the
equation of state derived in the framework of relativistic mean--field
theory. Both equations of state are rather similar, as can be seen in
Figs. \ref{fig:eos} and \ref{fig:mass}, and so are the corresponding
cooling curves, Fig. \ref{fig:g300_all_sf}. Again only the pion
condensation case without DU process is studied.

%%%%%%%%% Discussion %%%%
\section{Summary}

Two recalculated neutrino-emitting processes in dense neutron star
 matter, i.e., the modified Urca process and the superfluid pair
 breaking and formation processes, have been implemented in numerical
 cooling simulations of neutron stars.  Our major finding is that
 standard cooling processes supplemented with the medium effects
 enables one to achieve theoretical agreement with a large fraction of
 the observed pulsar luminosities, which --- subject to uncertainties
 in the equation of state and transport properties of superdense
 matter --- seems to be a problem for the standard cooling scenario
 alone (cf. Schaab et al. \cite{Schaab95a}). In particular it is
 possible to account for the low temperatures of PSRs 0833-45 (Vela),
 0656+14, and 0630+18 (Geminga).  Secondly, the new processes studied
 here provide a crossover between standard cooling and the so--called
 fast cooling scenarios of neutron stars and thus can be viewed as
 ``\emph{intermediate}'' cooling scenarios.

%%%%%%%%%%%%%%%%%%%%%%%%%%%%%%%%%%%%%%%%%%%%%%%%%%%%%%%%%%%%%%%%%%%%%%%%%%%%%
\begin{acknowledgements}
We would like to thank K. A. Van Riper for sending us his table of the
temperature gradient in the photosphere as a function of the surface
temperature, which was used in our cooling calculations.  We are also
grateful to the referee for his valuable comments.  Furthermore, we thank B. Friman and D. Page
for helpful discussions. D.V. thanks the GSI and Rostock University
for their hospitality and financial support and acknowledges financial
support from the International Science Foundation (grant N3W000).
\end{acknowledgements}

\end{document}